\documentclass[10pt]{article}
\usepackage{amsmath,amssymb,bm}
\usepackage[utf8]{inputenc}
\usepackage[english]{babel}
\usepackage{latexsym}
\usepackage{amsfonts}
\usepackage{mathtools}
\usepackage{amssymb}
\usepackage{scalerel}
\usepackage{extpfeil}
\usepackage[left=3cm,right=3cm,top=3.1cm,bottom=3.1cm]{geometry}
\usepackage{cite}
\usepackage{tensor}
\usepackage{tikz}
\usepackage{tikz-cd}

\usepackage[mathscr]{euscript}
\usetikzlibrary{arrows.meta, bending}
\tikzset{C/.style={circle, minimum size=8mm,
		node contents={},
		append after command={\pgfextra{%
				\draw[-{Straight Barb[flex']}](\tikzlastnode.150) arc (450:110:2.8mm);}
	}}
}

\usepackage{hyperref}
\numberwithin{equation}{section}

    \newcommand{\Tr}{\mathop{\rm Tr}\nolimits}

    \def\ket#1{|#1 \rangle}

    \def \be {\begin{eqnarray}}
    \def \ee {\end{eqnarray}}
    \def \bal {\begin{align}}
    \def \eal {\end{align}}
    \def \bdm {\begin{displaymath}}
    \def \edm {\end{displaymath}}

    \def\0{\nonumber}

    \def\wc{\omega_\text{c}}
    \def\wo{\omega_\text{o}}
    \def\hw{\hat\omega}

    \def\bfm{\boldsymbol{m}}
    \def\bfl{\boldsymbol{l}}
    \def\bfn{\boldsymbol{n}}
    
    \def \bfU{\boldsymbol{U}}

   \newcommand{\ch}[1]{{\color{black}{#1}}}

\begin{document}
	\begingroup\allowdisplaybreaks

\vspace*{1.1cm}

\centerline{\Large \bf Open-Closed String Field Theory }
\vspace {.3cm}

\centerline{\Large \bf in the Large $N$ Limit} 

\vspace{.3cm}

\begin{center}

{\large Carlo Maccaferri$^{(a)}$\footnote{Email: maccafer at gmail.com}, Alberto Ruffino$^{(a)}$\footnote{Email: ruffinoalb at gmail.com}  and  Jakub Vo\v{s}mera$^{(b)}$\footnote{Email: jvosmera at phys.ethz.ch} }
\vskip 1 cm
$^{(a)}${\it Dipartimento di Fisica, Universit\`a di Torino, \\INFN  Sezione di Torino \\
Via Pietro Giuria 1, I-10125 Torino, Italy}
\vskip .5 cm
\vskip .5 cm
$^{(b)}${\it Institut f\"{u}r Theoretische Physik, ETH Z\"{u}rich\\
	Wolfgang-Pauli-Strasse 27, 8093 Z\"{u}rich, Switzerland}

%
\end{center}

\vspace*{6.0ex}

\centerline{\bf Abstract}
\bigskip
We use the new nilpotent formulation of open-closed string field theory to explore the limit where the number $N$ of identical D-branes of the starting background is large. By reformulating the theory in terms of the 't Hooft coupling $\lambda\coloneqq\kappa N$, where $\kappa$ is the string coupling constant, we explicitly see that at large $N$ only genus zero vertices with arbitrary number of boundaries survive. After discussing the homotopy structure of the obtained large $N$ open-closed theory we discuss the possibility of integrating out  the open string sector with a quantum but planar homotopy transfer. As a result we end up with a classical closed string field theory, 
 described by a weak $L_\infty$-algebra containing a tree-level tadpole which, to first order in $\lambda$, is given by the initial boundary state. We discuss the possibility of removing the tadpole with a closed string vacuum shift solution, to end up with a new classical closed string background, where the initial D-branes have been turned into pure closed-string backreaction.
\baselineskip=16pt
\newpage
\tableofcontents

\section{Introduction}\label{sec:1}
Recently, covariant open-closed string field theory (OC-SFT), initially constructed in \cite{Zwiebach:1990qj, Zwiebach:1997fe}, has received a renewed attention as it has provided a complete and convenient framework where to address important D-brane effects in string theory. These include D-instanton contributions to closed string amplitudes  \cite{Chakravarty:2022cgj, Eniceicu:2022dru,
Eniceicu:2022nay, Alexandrov:2022mmy,  Agmon:2022vdj,
Alexandrov:2021dyl, Alexandrov:2021shf, Sen:2021jbr,
Sen:2021qdk, Sen:2020eck, Sen:2020ruy, Sen:2020oqr, Sen:2020cef, Sen:2019qqg} and  deformations of the  open string physics on a given D-brane system in a changing closed string background \cite{cosmo}. In particular in this last exploration, it has been possible to directly appreciate how the relevant observables describing the process of D-brane deformation are well-defined only in the framework of OC-SFT where it is  possible to generalize the `open string conjectures' (see  \cite{Erler:2019vhl}  for  a review) to the case where both the closed string background and the D-brane system change.  The gauge invariant observables describing the change in the background are associated to open-closed amplitudes and  the framework of OC-SFT, beside giving the most complete target space picture of the process, turns out to be just perfect in dealing  with IR divergences from both closed and open string degenerations.

At the same time however, this analysis made evident another fact. When we study how a change in the closed string background affects the D-brane physics, we are first of all solving the closed string field theory equation of motion to describe the change in the bulk CFT. However, since closed strings are coupled to the initial D-brane, we are not really solving the full string equation of motion but we are ignoring the backreaction that the D-brane has on the closed string background. In other words we are using the D-brane in probe approximation. A simple field theory analogue would be  to couple General Relativity (GR) to some matter,  in a background given by a vacuum GR solution, ignoring the gravitational backreaction of the matter. While this is a useful microscopic understanding of several defomations of gauge theories, we  would also like to have a microscopic understanding of the complementary problem: how does a given D-brane system affect the closed string geometry?  Roughly this should be captured by the sourced closed string equation of motion
\begin{align}
Q_{\rm c}\ket\Phi\sim \ch{\kappa}\ket {B_0},\label{source}
\end{align}
where $Q_{\rm c}$ is the closed string BRST charge, $\kappa$ is the closed string coupling constant and $\ket {B_0}$ is the boundary state of the D-brane system of which we want to capture the backreaction. However  this is only a very crude approximation to the backreaction because several  non-linearities enter the game due to the open-closed  and the closed-closed interactions. It takes a moment to realize that, differently from the GR-matter example (or, even simpler, a charge in interaction with an electromagnetic field) this problem is intrinsically quantum mechanical, since closed string propagation in presence of world-sheet boundaries is  indistinguishable from an open string loop, due to the open-closed channel duality. This problem \ch{sits} at the \ch{heart} of string theory and it has led  to the gauge/gravity duality (see  \cite{Aharony:1999ti} for review),  a profound conjecture whose proof is however still a challenge.
In \ch{the} absence of a full microscopic understanding in the \ch{critical-string setting}, this problem has been studied in specific simplified examples where the background is such that the local string dynamics is drastically simplified or absent, for example in the topological string \cite{Witten:1992fb, Gopakumar:1998ki, Ooguri:2002gx} or in the non-critical string \cite{Gaiotto:2003yb, McGreevy:2003kb, Douglas:2003up, Sen:2003iv} and also in some specific time-dependent backgrounds \cite{Gaiotto:2003rm, Lambert:2003zr} of critical string theory.  In these cases it has been indeed possible to reconstruct the full closed string backreaction from the open strings (or, more precisely, the D-branes) moduli. The typical  key move to access this has been to start from a large number $N$ of D-branes and then  focus on the 't Hooft limit $N\to\infty$ while keeping the 't Hooft coupling $\lambda=\kappa N$ fixed (or sending it to a critical value $\lambda_c$, \cite{ McGreevy:2003kb}). Since in this limit $\kappa$ is guaranteed to be very small, the world-sheet description is  valid and trustable. In this limit  the relevant string amplitudes are dominated by genus zero surfaces with an arbitrary number of boundaries, weighted by the 't Hooft coupling. Resumming the boundaries should then result in pure closed string amplitudes on the backreacted background as in the paradigmatic example of the Gopakumar-Vafa conifold transition \cite{Gopakumar:1998ki}. \\

It would be certainly desirable to have an understanding of this highly non-linear process in a complete space-time QFT approach such as  OC-SFT. \footnote{In fact this possibility has been already advocated in the early days \cite{Douglas:1996yp} before the $AdS/CFT$ breakthrough.}

In this paper  we propose a concrete description of D-branes' backreaction in OC-SFT, when this is formulated on a large number of D-branes.

Let us summarize our results by giving the plan of the paper. In section \ref{sec:2} we discuss how Zwiebach's OC-SFT looks like when we take as a starting background a large $N$ number of identical D-branes and we indeed verify that in the large $N$ limit all the couplings entering in the action are dominated by off-shell amplitudes on genus zero surfaces with an arbitrary number of boundaries.  In section \ref{sec:3} we study how the large $N$ limit affects the quantum homotopy relations which are at the basis of the perturbative consistency of OC-SFT. In doing this we take advantage of expressing the theory in the new nilpotent formulation discussed in our recent paper \cite{quantum-OC} which allows to write the full open-closed planar action in terms of a single open-closed planar coderivation $\bfn^{(p)}$ which packages at once all  open-closed couplings on genus zero Riemann surfaces with an arbitrary number of boundaries. 

  We explicitly verify that in the large $N$ limit of the BV quantum master equation  the closed string symplectic laplacian is suppressed as 
\begin{align}
\Delta_{\rm c}\sim \frac{\lambda^2}{N^2},
\end{align}
consistently with the fact that closed strings are classical in this limit. In addition also the open string BV laplacian gets simplified to the sole contribution where two open string punctures are glued together only if they sit on the same boundary,
\begin{align}
\Delta_{\rm o}{\Big |}_{\textrm {same boundary}}\coloneqq \Delta_{\rm o}^{(1)}\sim \lambda,
\end{align}
while  the gluing of open punctures from different boundaries is suppressed
\begin{align}
\Delta_{\rm o}{\Big |}_{\textrm {different boundaries}}\coloneqq \Delta_{\rm o}^{(2)}\sim \frac\lambda{N^2}.
\end{align}
The planar BV master equation 
\begin{align}
(S,S)_{\rm c}+(S,S)_{\rm o}+2\Delta_{\rm o}^{(1)} S=0
\end{align}
then translates into the planar nilpotency relation
\begin{align}
(\bfn^{(p)}+\bfU^{(p)})^2=0,
\end{align}
where $\bfU^{(p)}$ is the planar open string Poisson bi-vector corresponding to $\Delta_{\rm o}^{(1)}$.
In section \ref{sec:4.0} we address the possibility of capturing the closed string backreaction of the large number of branes we have started with. In \ref{sec:4} we discuss the  integration  out of the open string sector using the homotopy transfer
 \cite{Sen:2016qap, Jurco:2019yfd, Erbin:2020eyc, Koyama:2020qfb, Okawa:2022sjf } on the planar open-closed coderivation $\bfn^{(p)}$.  We observe that the structure of the homological perturbation lemma remains consistent if we can assume that the open string cohomology doesn't flow through the internal open string propagators. 
 Under this assumption we derive the final closed string effective couplings which can be identified with off-shell closed strings amplitudes on genus zero surfaces with arbitrary number of boundaries, with a full moduli space integration towards open string degeneration. These couplings are finite and well-defined as long as it is consistent to ignore the open string cohomology as  above. We show that these couplings define a weak $L_\infty$ algebra, meaning that the obtained closed string field theory action is purely classical but it  has a non-vanishing  tadpole.  Then in \ref{sec:5} we discuss the elimination of the tadpole through a closed string vacuum shift solution $\Phi(\lambda)$ whose equation of motion reads 
\begin{align}
\sum_{k=1}^\infty\,\frac1{k!}\tilde l_{k}(\Phi(\lambda)^{\wedge k})+\tilde l_0=0\sim\lambda\left(Q_{\rm c}\Phi'(0)+\ch{\ket{B_0'}}\right)+O(\lambda^2)\,,
\end{align} 
\ch{where the normalized boundary state $\ket{B_0'} = (1/N)\ket{B_0}$ has a finite large $N$ limit and coincides with the boundary state of a single D-brane of the $N$-stack.} \ch{This} provides the searched-for non-linear completion of \eqref{source}. If a regular solution to this equation exists, the closed string theory will cancel the tadpole and it will thus describe a new purely closed string background which is the result of the backreaction of the initial $N$ D-branes. We conclude in \ref{sec:6} with a final discussion. Appendix \ref{app:A} contains technical results on the uplift of the propagator and the other objects entering the strong deformation retract at the  basis of the open string integration-out, to the open-closed tensor algebra.

\section{Open-Closed string field theory in the large $N$ limit}\label{sec:2}
 Consider the BV quantum master action of bosonic open-closed string field theory \cite{Zwiebach:1997fe, quantum-OC}
  \begin{equation}
        S_{\rm oc}[\Phi,\Psi]=\sum_{g=0}^{\infty}\sum_{b=0}^{\infty}\kappa^{2g+b-2}\sum_{k=0}^{\infty}\sum_{\{l_{1},...,l_{b}\}=0}^{\infty}\frac{1}{b!k!(l_1)\cdots(l_b)}\mathcal{A}^{g,b}_{k;\{l_{1},...,l_{b}\}}\left(\Phi^{\wedge k}\otimes'\Psi^{\odot l_{1}}\wedge'...\wedge' \Psi^{\odot l_{b}}\right),\label{OC-action}   
    \end{equation}
where $\Phi$ is the closed BV string field, $\Psi$ is the open one and $\kappa$ is the string coupling constant.  See \cite{quantum-OC} for our conventions. In particular $\odot$ is the cyclic tensor product, which is used to place color-ordered open strings on every boundary.
The off-shell amplitudes $\mathcal{A}^{g,b}_{k;\{l_{1},...,l_{b}\}}$ are computed in a given BCFT$_{0}$ which describes the initial open-closed background, represented in the closed string channel by a given boundary state $\ket {B'_0}$. 

Now we want to consider this theory  defined on $N$ identical BCFT$_0$ D-branes, described by the boundary state $N\ket {B'_0}$.
This means to add $N\times N$ Chan-Paton factors to the open string field $\Psi$. In particular if the original BV open string field was expanded as $\Psi=\psi^a\,o_a$ (where $o_a$ is a basis of BCFT$_0$ boundary fields and $\psi^a$ are collectively the open BV fields and anti-fields), now we will have
\begin{align}
\Psi=\psi^{a}_{ij}\,t^{ij}o_a,\label{open-chan}
\end{align}
where $\{t^{ij}\}$ is a basis for $N\times N$ matrices which for definiteness we take as
\begin{align}
(t^{ij})_{pq}=\delta^i_p\,\delta^j_q.\label{t-comp}
\end{align}
In doing so we would like to re-write the action in terms of objects which remain finite as the limit $N\to\infty$ is taken. 
To do so it is convenient to define the normalized trace for $N\times N$ matrices
\begin{align}
\Tr'[M]\coloneqq \frac1N\,\Tr[M],
\end{align}
in such a way that, for example, $\Tr'[\boldsymbol{1}_N]=1$.
The off-shell amplitude $\mathcal{A}^{g,b}$  contains a number of traces over CP factors which equals the number of boundaries. We can thus define normalized off-shell amplitudes as
\begin{align}
\mathcal{A'}^{g,b}\coloneqq\frac1{N^{b}}\,\mathcal{A}^{g,b},
\end{align}
which remain finite when $N\to\infty$. If we write the action \eqref{OC-action} in terms of normalized CP traces we thus get
\begin{align}
   S_{\rm oc}[\Phi,\Psi]=&\sum_{g=0}^{\infty}\sum_{b=0}^{\infty}\kappa^{2g+b-2}\,N^b\,\sum_{k=0}^{\infty}\sum_{\{l_{1},...,l_{b}\}=0}^{\infty}\frac{1}{b!k!(l_1)\cdots(l_b)}\mathcal{A'}^{g,b}_{k;\{l_{1},...,l_{b}\}}\left(\Phi^{\wedge k}\otimes'\Psi^{\odot l_{1}}\wedge'...\wedge' \Psi^{\odot l_{b}}\right)\0\\
  =&\sum_{g=0}^{\infty} \kappa^{2g-2}\sum_{b=0}^{\infty} \lambda^b\sum_{k=0}^{\infty}\sum_{\{l_{1},...,l_{b}\}=0}^{\infty}\frac{1}{b!k!(l_1)\cdots(l_b)}\mathcal{A'}^{g,b}_{k;\{l_{1},...,l_{b}\}}\left(\Phi^{\wedge k}\otimes'\Psi^{\odot l_{1}}\wedge'...\wedge' \Psi^{\odot l_{b}}\right),
  \label{OC-kappa-lambda-action}  
\end{align}
where we have defined the 't Hooft coupling
\begin{align}
\lambda\coloneqq \kappa N.
\end{align}
Alternatively we can also write
\begin{align}
 S_{\rm oc}[\Phi,\Psi]=\sum_{g=0}^{\infty}\sum_{b=0}^{\infty}N^{2-2g}\lambda^{2g+b-2}\sum_{k=0}^{\infty}\sum_{\{l_{1},...,l_{b}\}=0}^{\infty}\frac{1}{b!k!(l_1)\cdots(l_b)}\mathcal{A'}^{g,b}_{k;\{l_{1},...,l_{b}\}}\left(\Phi^{\wedge k}\otimes'\Psi^{\odot l_{1}}\wedge'...\wedge' \Psi^{\odot l_{b}}\right).
 \label{OC-N-lambda-action}  
\end{align}
We are interested in the open-closed  action in the large $N$ limit at fixed 't Hooft coupling $\lambda$. In this limit we notice that only genus zero surfaces survive
\begin{align}
\lim_{N\to\infty}\frac1{N^2}S_{\rm oc}[\Phi,\Psi]=&\sum_{b=0}^{\infty} \lambda^{b-2}\sum_{k=0}^{\infty}\sum_{\{l_{1},...,l_{b}\}=0}^{\infty}\frac{1}{b!k!(l_1)\cdots(l_b)}\mathcal{A'}^{g=0,b}_{k;\{l_{1},...,l_{b}\}}\left(\Phi^{\wedge k}\otimes'\Psi^{\odot l_{1}}\wedge'...\wedge' \Psi^{\odot l_{b}}\right),\0\\
\coloneqq&S_{\rm pl}[\Phi,\Psi],\label{S-planar}
\end{align}
where ``pl'' stands for {\it planar}.
In this paper we will be interested in exploring  this limit. 
 
 \section{Large $N$ homotopy algebra}\label{sec:3}
In this section we will reformulate what we have just said  using the open-closed tensor algebra notation introduced in \cite{quantum-OC} and we will obtain the large-$N$ open-closed homotopy relations from the quantum BV master equation. Since, if not for the $N$ dependence, this is  a rewriting of what has already been done in detail in \cite{quantum-OC}, we will often just give the results and only comment on the instances where the $N$ parameter plays an important role.

\subsection{Normalized multistring products}

To start with we rewrite the action using multistring products.
First of all the normalized amplitudes involving only closed strings are written as
\begin{align}
\mathcal{A'}^{g,b}_{k+1}(\Phi_1\wedge\cdots\wedge\Phi_{k+1})\coloneqq\wc\left(\Phi_1,{l'}^{(g,b)}_{k,0}\left(\Phi_2\wedge\cdots\wedge \Phi_{k+1}\right)\right),
\end{align}
where $\wc$ is the closed string symplectic form (used in the  standard conventions of \cite{quantum-OC}) and we have defined the normalized multistring products
\begin{align}
 {l'}^{(g,b)}_{k,0}:&\,{\cal H}_{\rm c}^{\wedge k}\,\rightarrow\,{\cal H}_{\rm c},\\
{l'}^{(g,b)}&\coloneqq\,N^{-b}\,l^{(g,b)},
  \end{align}
  where $l^{(g,b)}$ is the non-normalized product (that is with usual CP traces on every boundary).
The ${l'}^{(g,b)}$  are cyclic with respect to $\wc$
    \begin{align}
    \wc\left(\Phi_1,{l'}^{(g,b)}_{k,0}\left(\Phi_2\wedge\cdots\wedge \Phi_{k+1}\right)\right)=(-1)^\epsilon \wc\left(\Phi_{k+1},{l'}^{(g,b)}_{k,0}\left(\Phi_1\wedge\cdots\wedge \Phi_{k}\right)\right)
    \end{align}
    and  BPZ odd
    \begin{align}
    \wc\left(\Phi_1,{l'}^{(g,b)}_{k,0}\left(\Phi_2\wedge\cdots\wedge \Phi_{k+1}\right)\right)=-(-1)^{d(\Phi_1)} \ch{\wc}\left({l'}^{(g,b)}_{k,0}\left(\Phi_1\wedge\cdots\wedge \Phi_{k}\right),\Phi_{k+1}\right).\label{bpz-cycl-closed}
    \end{align}
The other normalized amplitudes containing at least one open string, can be written using the open string symplectic form,  choosing a special boundary, as explained in detail in \cite{quantum-OC}
    \begin{align}
    &\mathcal{A'}^{g,b}_{k;\{l_{1},...,l_{b}\}}\left(\Phi_1\wedge\cdots\wedge\Phi_k\otimes'\Psi_{1,1}\odot\cdots\odot\Psi_{1, l_{1}}\,\wedge'...\wedge' \Psi_{b,1}\odot\cdots\odot\Psi_{b, l_{b}}\right)\nonumber\\
    \coloneqq&\wo'\left(\Psi_{b,1}, {m'}^{(g,b)}_{k\left[l_{1},...,l_{b-1}\right]l_{b}-1}\left((\Phi)_k\otimes' [\Psi]_{ l_{1}}\wedge' \cdots\wedge'  [\Psi]_{ l_{b-1}}\otimes''\Psi_{b,2}\otimes\cdots\otimes\Psi_{b,l_b}\right)\right),
    \end{align}
    where 
    \begin{align}
    (\Phi)_k&\coloneqq\Phi_1\wedge\cdots\wedge\Phi_k\0\\
    [\Psi]_{l_j}&\coloneqq\Psi_{j,1}\odot\cdots\odot\Psi_{j, l_{j}}\,.
    \end{align}
  The normalized open string symplectic form is defined with a normalized Chan-Paton trace 
 \begin{align}
 \wo'\coloneqq \frac{\wo}{N}.
 \end{align}
 The normalized multistring products with open string output ${m'}^{(g,b)}$ are odd linear maps
    \begin{align}\label{mdef}
    {m'}^{(g,b)}_{k\left[l_{1},...,l_{b-1}\right]l_{b}}:& \,{\cal H}_{\rm c}^{\wedge k}\otimes'{\cal H}_{\rm o}^{\odot l_1}\wedge'\cdots\wedge'{\cal H}_{\rm o}^{\odot l_{b-1}}\otimes''{\cal H}_{\rm o}^{\otimes l_b}\rightarrow\,{\cal H}_{\rm o},\\
    {m'}^{(g,b)}\coloneqq& N^{1-b}\,m^{(g,b)},
    \end{align}
     which are cyclic  
    \begin{align}
    &\wo'\left(\Psi_{b,1}, {m'}^{(g,b)}_{k\left[l_{1},...,l_{b-1}\right]l_{b}-1}\left((\Phi)_k\otimes' [\Psi]_{ l_{1}}\wedge' \cdots\wedge'  [\Psi]_{ l_{b-1}}\otimes'' \Psi_{b,2}\otimes\cdots\otimes\Psi_{b,l_b}\right)\right)\\
    =&(-1)^\epsilon\wo'\left(\Psi_{b,l_b}, {m'}^{(g,b)}_{k\left[l_{1},...,l_{b-1}\right]l_{b}-1}\left((\Phi)_k\otimes' [\Psi]_{ l_{1}}\wedge' \cdots\wedge'  [\Psi]_{ l_{b-1}}\otimes'' \Psi_{b,1}\otimes\cdots\otimes\Psi_{b,l_{b}-1}\right)\right)\0
    \end{align}
    and BPZ odd 
    \begin{align}
    &\wo'\left(\Psi_{b,1}, {m'}^{(g,b)}_{k\left[l_{1},...,l_{b-1}\right]l_{b}-1}\left((\Phi)_k\otimes' [\Psi]_{ l_{1}}\wedge' \cdots\wedge'  [\Psi]_{ l_{b-1}}\otimes'' \Psi_{b,2}\otimes\cdots\otimes\Psi_{b,l_b}\right)\right)\label{open-cycl1}\\
    =&-(-1)^\epsilon\wo'\left( {m'}^{(g,b)}_{k\left[l_{1},...,l_{b-1}\right]l_{b}-1}\left((\Phi)_k\otimes' [\Psi]_{ l_{1}}\wedge' \cdots\wedge'  [\Psi]_{ l_{b-1}}\otimes'' \Psi_{b,1}\otimes\cdots\otimes\Psi_{b,l_{b}-1}\right),\Psi_{b,l_b}\right)\0.
    \end{align}
    Precisely as in the non-normalized case \cite{quantum-OC}, there is another \ch{cyclicity} relation obtained by changing the choice of special boundary
    \begin{align}
    &\wo'\left(\Psi_{b,1}, {m'}^{(g,b)}_{k\left[l_{1},...,l_{b-1}\right]l_{b}-1}\left((\Phi)_k\otimes' [\Psi]_{ l_{1}}\wedge' \cdots\wedge'  [\Psi]_{ l_{b-1}}\otimes'' \Psi_{b,2}\otimes\cdots\otimes\Psi_{b,l_b}\right)\right)\0\\
    =&(-1)^\epsilon\,\wo'\left(\Psi_{1,1}, {m'}^{(g,b)}_{k\left[l_{2},...,l_{b}\right]l_{1}-1}\left((\Phi)_k\otimes' [\Psi]_{ l_{2}}\wedge' \cdots\wedge'  [\Psi]_{ l_{b}}\otimes''\Psi_{1,2}\otimes\cdots\otimes\Psi_{1,l_1}\right)\right)\label{open-cycl2}\\
    =&\mathcal{A'}^{g,b}_{k;\{l_{1},...,l_{b}\}}\left((\Phi)_k\otimes'[\Psi]_{l_1}\wedge'...\wedge' [\Psi]_{l_b}\right)\0.
    \end{align}
We then define \ch{(non-independent)} normalized  products ${l'}^{(g,b)}$ with a closed string output such that the same normalized open-closed amplitude can be computed with the normalized-open or closed symplectic form 
    \begin{align}
    &\wo'\left(\Psi_{b,1}, {m'}^{(g,b)}_{k\left[l_{1},...,l_{b-1}\right]l_{b}-1}\left((\Phi)_k\otimes' [\Psi]_{ l_{1}}\wedge' \cdots\wedge'  [\Psi]_{ l_{b-1}}\otimes'' \Psi_{b,2}\otimes\cdots\otimes\Psi_{b,l_b}\right)\right)\0\\
    \coloneqq&(-1)^\epsilon \wc\left(\Phi_1,{l'}^{(g,b)}_{k-1,[l_1,\cdots, l_b]}\left(\Phi_2\wedge\cdots\wedge\Phi_k\otimes'[\Psi]_{l_1}\wedge'\cdots\wedge'[\Psi]_{l_b}\right)\right)\label{dual}\\
    =&\mathcal{A'}^{g,b}_{k;\{l_{1},...,l_{b}\}}\left((\Phi)_k\otimes'[\Psi]_{l_1}\wedge'...\wedge' [\Psi]_{l_b}\right)\0.
    \end{align}
    With the definitions and properties given above, the open-closed BV action \eqref{OC-action} can be written as
      \begin{align}
    &S_{\rm oc}[\Phi,\Psi]=\ch{\Lambda_0+}\sum_{g,b=0}^{\infty}\kappa^{2g-2}\,\frac{\lambda^b}{b!}\,\sum_{k=0}^{\infty}{{\Bigg [}}\frac1{(k+1)!}
    \wc\left(\Phi,l'^{(g,b)}_{k,0}\left(\Phi^{\wedge k}\right)\right)\label{OCact}\\
    &+{\sum_{\substack{\{l_{1},...,l_{b-1}\}=0\\ {l_b=1}}}^{\infty}}\frac{{C({l_1,\ldots, l_b})}}{k!\,(l_1)\cdots(l_b)}
    \wo'\left(\Psi,m'^{(g,b)}_{k\left[l_{1},...,l_{b-1}\right]l_{b}-1}\left(\Phi^{\wedge k}\otimes'\Psi^{\odot l_1}\cdots\Psi^{\odot l_{b-1}}\otimes''\Psi^{\otimes (l_b-1)}\right)\right){{\Bigg ]}},\0
    \end{align}
     where the normalizations $(l)$ due to cyclicity are defined by $(l)\coloneqq l+\delta_{l,0}$. 
   The combinatorial factor $C({l_1,\ldots, l_b})$ accounts for the choice of the special boundary $b$ (on which we compute the normalized open string symplectic form $\wo'$) in the set of non-empty boundaries, and it is defined and discussed in \cite{quantum-OC}.  This combinatorial factor will be automatically accounted for in the WZW-like co-algebraic formulation in the next section.
     The constant $\Lambda_0$ (which we did not write explicitly in \cite{quantum-OC} as it is not important for the local open-closed dynamics) is a constant of the initial microscopic action formally given by
 \begin{align}
 \Lambda_0&=\sum_{g=0}^{\infty} \kappa^{2g-2}\sum_{b=0}^{\infty} \frac{\lambda^b}{b!}\,\mathcal{A'}^{g,b}_{0;0,...,0}.\label{Lambda0}
 \end{align}
 As the notation suggests it corresponds to the world-sheet partition function summed over boundaries and genera, but cut-off at closed and open string degenerations.  We expect that, for $g=0$,  its value should be such that  when the open-closed action is expanded around the saddle point of the open-string tachyon vacuum, the overall constant term \ch{should become equal to the sphere partition function}, see \cite{cosmo} for the classical aspects of this.

    In this form of the action, we will be interested in the $\kappa\to0$ limit, with $\lambda$ constant. Equivalently we can also consider
    \begin{align}
    \kappa^{2g-2}\lambda^b=\frac1{  N^{2g-2}}\lambda^{2g+b-2},
    \end{align}
    and take the $N\to\infty$ limit, always with $\lambda$ fixed.

\subsection{Normalized coderivations}

The upgrade of the above-defined normalized products to normalized coderivations is identical to what we did in \cite{quantum-OC} so we just give the  results. Starting from the normalized products with closed string output ${l'}^{(g,b)}_{k\left[l_{1},...,l_{b}\right]}$ we first upgrade them to coderivations ${\bfl'}^{(g,b)}_{k\left[l_{1},...,l_{b}\right]}$ and then we define the total closed string coderivation at fixed $(g,b)$
\begin{align}
    {\bfl'}^{(g,b)}\coloneqq\sum_{k,\left\{l_1,...,l_b\right\}} {\bfl'}^{(g,b)}_{k\left[l_{1},...,l_{b}\right]}.
    \end{align}
Similarly for the products with open string output ${m'}^{(g,b)}$ we define
\begin{align}
{\bfm'}^{(g,b)}\coloneqq\sum_{k\geq0}\sum_{l_1\cdots l_b\geq0}{\bfm'}^{(g,b)}_{k\left[l_{1},...,l_{b-1}\right]l_{b}}.
\end{align} 
Just as their non-normalized versions of \cite{quantum-OC} these normalized coderivations obey the the cyclicity relations
 \begin{align}
    \wc\left(\pi_{10}\, {\bfl'}^{(g,b)}\,\boldsymbol{a}_{\rm c}\,{\cal G}\,,\,\pi_{10}\boldsymbol{b}_{\rm c}\,{\cal G}\right)&=-(-1)^{d(a)}\wc\left(\pi_{10} \,\boldsymbol{a}_{\rm c}\,{\cal G}\,,\,\pi_{10}{\bfl'}^{(g,b)}\boldsymbol{b}_{\rm c}\,{\cal G}\right),\label{clcycl}\\
      \wo'\left(\pi_{01} {\bfm'}^{(g,b)}\boldsymbol{a}_{\rm o}\,{\cal G},\pi_{01}\boldsymbol{b}_{\rm o}\,{\cal G}\right)&=-(-1)^{d(a)}\wo'\left(\pi_{01} \boldsymbol{a}_{\rm o}\,{\cal G},\pi_{01}{\bfm'}^{(g,b)}\boldsymbol{b}_{\rm o}\,{\cal G}\right), \label{opcycl}\\
\wc\left(\pi_{10}\,{\bfl'}^{(g,b)}\,\boldsymbol{a}_{\rm o}\,{\cal G}\,,\, \pi_{10}\,\boldsymbol{b}_{\rm c}\,{\cal G}\right)&=-(-1)^{d(a)}\wo'\left(\pi_{01}\,\boldsymbol{a}_{\rm o}\,{\cal G}\,,\, \pi_{01}\,{\bfm'}^{(g,b)}\,\boldsymbol{b}_{\rm c}\,{\cal G}\right) \label{opclcycl1} \\
\wo'\left(\pi_{01}\,{\bfm'}^{(g,b)}\,\boldsymbol{a}_{\rm c}\,{\cal G}\,,\, \pi_{01}\,\boldsymbol{b}_{\rm o}\,{\cal G}\right)&=-(-1)^{d(a)}\wc\left(\pi_{10}\,\boldsymbol{a}_{\rm c}\,{\cal G}\,,\, \pi_{10}\,{\bfl'}^{(g,b)}\,\boldsymbol{b}_{\rm o}\,{\cal G}\right),\label{opclcycl2}
    \end{align}
where $(\pi_{10},\pi_{01})$ project into a single copy of respectively $({\cal H}_{\rm c},{\cal H}_{\rm o})$ inside $\mathcal{SH}_{\rm c}\otimes' \mathcal{SCH}_{\rm o}$ and $(\boldsymbol{a}_{\rm c/o},\boldsymbol{b}_{\rm c/o})$ are any two closed or open string coderivations ({\it i.e.} with closed or  open output respectively). All of these relations are just normalized re-writings of the corresponding relations of \cite{quantum-OC}. In particular, to help the reader, we recall that ${\cal G}$ is the open-closed group like element
\begin{align}
    {\cal G}\coloneqq&e^{\wedge \Phi}\otimes'\,e^{\wedge' {\cal C}(\Psi)}\0\\
    =&\left[\sum_{k\geq0}\frac1{k!} \Phi^{\wedge k}\right]\otimes'\left[\sum_{b\geq0}\frac1{b!}\left(\sum_{l_1\geq0}\frac{1}{(l_1)}\Psi^{\odot l_1}\right)\wedge'\cdots\wedge'\left(\sum_{l_b\geq0}\frac1{(l_b)}\Psi^{\odot l_b}\right)\right],
    \end{align}
where closed strings are fully symmetrized, open strings are cyclic on every boundary  and  all boundaries are symmetrized between themselves.

Going on,  we sum up the open and closed coderivations as
\begin{align}
\bfl\coloneqq \sum_{g,b}\kappa^{2g+b}\bfl^{(g,b)}=& \sum_{g,b}\kappa^{2g+b}N^b\,{\bfl'}^{(g,b)} =\sum_{g,b}\kappa^{2g}\lambda^b \,{\bfl'}^{(g,b)}\label{l-expand}\\
\bfm\coloneqq \sum_{g,b}\kappa^{2g+b-1}\bfm^{(g,b)}=& \sum_{g,b}\kappa^{2g+b-1}N^{b-1}\,{\bfm'}^{(g,b)} =\sum_{g,b}\kappa^{2g}\lambda^{b-1} \,{\bfm'}^{(g,b)}.\label{m-expand}
 \end{align}
Notice that the resummed $\bfl$ and $\bfm$ are the same as in the un-normalized case treated in \cite{quantum-OC}, but this time we express them using normalized coderivations. Then we can write the full dynamical part of the action in WZW form 
\begin{align}
S_{\rm oc}(\Phi,\Psi)=&\int_0^1dt\left(\frac{\wc}{\kappa^2}\left(\dot \Phi\,,\,\pi_{10} \bfl\,{\cal G}\right)+ \frac\wo\kappa\left(\dot \Psi\,,\,\pi_{01}\bfm\,{\cal G}\right)\right)\0,\\
=&N^2\int_0^1dt\left(\frac{\wc}{\lambda^2}\left(\dot \Phi\,,\,\pi_{10} \bfl\,{\cal G}\right)+ \frac{\wo'}\lambda\left(\dot \Psi\,,\,\pi_{01}\bfm\,{\cal G}\right)\right),\label{OC-WZWoc}
\end{align}
where in the second line we have switched to the normalized $\wo'$ and we have have used $\lambda=\kappa N$. In the 't Hooft limit $N\to\infty$ with $\lambda$ constant the planar action \eqref{S-planar} is thus written as
\begin{align}
S_{\rm pl}(\Phi,\Psi)=\lim_{N\to \infty}\frac1{N^2} S_{\rm oc}(\Phi,\Psi)=\int_0^1dt\left(\frac{\wc}{\lambda^2}\left(\dot \Phi\,,\,\pi_{10} \bfl^{(p)}\,{\cal G}\right)+ \frac{\wo'}\lambda\left(\dot \Psi\,,\,\pi_{01}\bfm^{(p)}\,{\cal G}\right)\right),\label{OC-WZW-planar}
\end{align}
where we have defined the planar coderivations  $\bfl^{(p)}$ and  $\bfm^{(p)}$ as the genus zero contributions of $\bfl$ and $\bfm$.
\begin{align}
\bfl^{(p)}\coloneqq& \sum_b \lambda^b\,{\bfl'}^{(0,b)},\\
\bfm^{(p)}\coloneqq& \sum_b \lambda^{b-1}\,\ch{{\bfm'}^{(0,b)}}.
\end{align}
Notice that, like $\wo'$, these objects are finite in the 't Hooft limit, contrary to the higher genus contributions which are suppressed by $\kappa^{2g}\sim N^{-2g}$.

\subsection{Large $N$ BV master equation}

Following the conventions of \cite{cosmo} and \cite{quantum-OC} we write the quantum BV master equation as
\begin{equation}
\label{eqn: BVquantummasterequation}
    \dfrac{1}{2}\left(S_{\rm oc},S_{\rm oc}\right)+\Delta S_{\rm oc}=0,
\end{equation}
where we have set $\hbar=1$. The BV brackets can be split into an open part and a closed part
\begin{equation}
\left(\cdot,\cdot\right)=\left(\cdot,\cdot\right)_{\rm c}+\left(\cdot,\cdot\right)_{\rm o}.
\end{equation}
To define these objects we need to recall how the closed and open BV string fields are decomposed. For the BV closed string field we have
\begin{align}
\Phi=\phi^a&\,c_a\\
\wc(c_a,c_b)\coloneqq &(\wc)_{ab}\\
\wc(c^a,c_b)\coloneqq & \delta^a_b\\
\wc(c^a,c^b)\coloneqq& (\wc)^{ab},
\end{align}
where $\phi^a$ are the BV fields and $(c_a, c^b)$ are respectively a basis and a dual basis for ${\cal H}_{\rm c}$. See \cite{quantum-OC} for further details. The open BV string field is decomposed analogously, paying attention that its basis now contain $N\times N$ Chan-Paton's factors \eqref{open-chan}
\begin{align}
\Psi=\psi_{ij}^a\,t^{ij}\,&o_a\\
\wo(t^{ij}o_a,t^{pq}o_b)\coloneqq \,&(\wo)_{ab}^{ij,pq}\\
\wo(t_{ij}o^a,t^{pq}o_b)\coloneqq \,& \delta^a_b\,\delta^p_i\,\delta^q_j\\
\wo(t_{ij}o^a,t_{pq}o^b)\coloneqq\,& (\wo)^{ab}_{ij,pq},
\end{align}
where we recall \eqref{t-comp} and we have defined the dual matrices 
\begin{align}
t_{ij}\coloneqq t^{ji}.
\end{align}
Let us now proceed by writing BV brackets in terms of normalized objects
\begin{equation}
    \left(\cdot, \cdot\right)_{\rm c}=\kappa^2\dfrac{\overleftarrow{\partial}}{\partial\phi^{a}}\omega_{\rm c}^{ab}\dfrac{\overrightarrow{\partial}}{\partial \phi^{b}}=\lambda^{2} N^{-2} \dfrac{\overleftarrow{\partial}}{\partial\phi^{a}}\omega_{\rm c}^{ab}\dfrac{\overrightarrow{\partial}}{\partial \phi^{b}},
\end{equation}
\begin{equation}
    \left(\cdot, \cdot\right)_{\rm o}=\kappa \dfrac{\overleftarrow{\partial}}{\partial\psi^{a}_{ij}}(\omega_{\rm o})^{ab}_{ij,kl}\dfrac{\overrightarrow{\partial}}{\partial \psi^{b}_{kl}}
=\lambda N^{-2} \dfrac{\overleftarrow{\partial}}{\partial\psi^{a}_{ij}}(\omega_{\rm o}')^{ab}_{ij,kl}\dfrac{\overrightarrow{\partial}}{\partial \psi^{b}_{kl}}.
\end{equation}
Similarly the symplectic laplacians are written as

\begin{equation}
    \Delta_{\rm c}=\kappa^2(-)^{\phi^{a}}\omega_{\rm c}^{ab}\dfrac{\overrightarrow{\partial}}{\partial \phi^{a}}\dfrac{\overrightarrow{\partial}}{\partial \phi^{b}}=
\dfrac{\lambda^{2} N^{-2}}{2}(-)^{\phi^{a}}\omega_{\rm c}^{ab}\dfrac{\overrightarrow{\partial}}{\partial \phi^{a}}\dfrac{\overrightarrow{\partial}}{\partial \phi^{b}},
\end{equation}
\begin{equation}
    \Delta_{\rm o}=\kappa(-)^{\psi^{a}_{ij}}(\omega_{\rm o})^{ab}_{ij,kl}\dfrac{\overrightarrow{\partial}}{\partial \psi^{a}_{ij}}\dfrac{\overrightarrow{\partial}}{\partial \psi^{b}_{kl}}=\dfrac{\lambda N^{-2}}{2}(-)^{\psi^{a}_{ij}}(\omega_{\rm o}')^{ab}_{ij,kl}\dfrac{\overrightarrow{\partial}}{\partial \psi^{a}_{ij}}\dfrac{\overrightarrow{\partial}}{\partial \psi^{b}_{kl}}.
\end{equation}
In the large $N$ limit it is important to split the open string laplacian  $\Delta_{\rm o}$ into two contributions where the two derivatives pick open strings from the same or from different boundaries
\begin{align}
\Delta_{\rm o}&=\Delta^{(1)}_{\rm o}+\Delta^{(2)}_{\rm o},
\end{align}
where 
\begin{align}
\Delta^{(1)}_{\rm o}&=\dfrac{\lambda N^{-2}}{2}(-)^{\psi^{a}_{ij}}(\omega_{\rm o}')^{ab}_{ij,kl}\dfrac{\overrightarrow{\partial}}{\partial \psi^{a}_{ij}}\dfrac{\overrightarrow{\partial}}{\partial \psi^{b}_{kl}}{\bigg |}_{\textrm{same boundary}}\\
\Delta^{(2)}_{\rm o}&=\dfrac{\lambda N^{-2}}{2}(-)^{\psi^{a}_{ij}}(\omega_{\rm o}')^{ab}_{ij,kl}\dfrac{\overrightarrow{\partial}}{\partial \psi^{a}_{ij}}\dfrac{\overrightarrow{\partial}}{\partial \psi^{b}_{kl}}{\bigg |}_{\textrm{different boundaries}}.
\end{align}
Geometrically  $\Delta_{\rm o}^{(1)}$ glues two punctures on the same  boundary thus adding one boundary to the surface. On the other hand  $\Delta_{\rm o}^{(2)}$ glues two punctures from different boundaries, thus reducing by one the number of boundaries but increasing the genus by one. 
These geometrical operations are reflected on the Chan-Paton matrices by the following trace identities which are easy to verify
\begin{equation}
\Delta^{(1)}_{\rm o}:\quad\quad    {\rm Tr}\left(t_{ij}At^{ij}B\right)={\rm Tr}(A){\rm Tr}(B),
\end{equation}
\begin{equation}
 \Delta^{(2)}_{\rm o}:\quad\quad   {\rm Tr}\left(t_{ij}A\right){\rm Tr}\left(t^{ij}B\right)={\rm Tr}(AB),
\end{equation}
which we write also using normalized traces
\begin{equation}
 \Delta^{(1)}_{\rm o}:\quad\quad   \dfrac{1}{N}{\rm Tr'}\left(t_{ij}At^{ij}B\right)={\rm Tr'}(A){\rm Tr'}(B),\label{ntrid1}
\end{equation}
\begin{equation}
\Delta^{(2)}_{\rm o}:\quad\quad  N{\rm Tr'}\left(t_{ij}A\right){\rm Tr'}\left(t^{ij}B\right)={\rm Tr'}(AB).\label{ntrid2}
\end{equation}
In total both $\Delta_{\rm o}^{(1,2)}$ have the  effect of decreasing the Euler number by one but topologically they are very different operations, which are finally  distinguished when the single parameter $\kappa$ is resolved to $\kappa$ and $\lambda$, thanks to the new parameter $N$. 

With these preparation we can now read-off from \cite{quantum-OC} the following results for the open-closed action \eqref{OC-WZWoc}
\begin{align}
\frac12(S_{\rm oc},S_{\rm oc})_{\rm c}=&N^2\left(\int_{0}^{1}dt\,\frac\wc{\lambda^2}\left(\dot\Phi,\pi_{10}\,\boldsymbol{l}\boldsymbol{l}\,{\cal G}\right)+\int_{0}^{1}dt\,\frac{\wo'}{\lambda}\left(\dot\Psi,\pi_{01}\boldsymbol{m}\boldsymbol{l}\,{\cal G}\right)\right),\\
\frac12(S_{\rm oc},S_{\rm oc})_{\rm o}=&N^2\left(\int_{0}^{1}dt\,\frac\wc{\lambda^2}\left(\dot\Phi,\pi_{10}\,\boldsymbol{l}\boldsymbol{m}\,{\cal G}\right)+\int_{0}^{1}dt\,\frac{\wo'}{\lambda}\left(\dot\Psi,\pi_{01}\boldsymbol{m}\boldsymbol{m}\,{\cal G}\right)\right),\\
\Delta_{\rm c}S_{\rm oc}=&N^2\left(\int_{0}^{1}dt\,\frac\wc{\lambda^2}\left(\dot\Phi,\pi_{10}\,\frac{\lambda^2} {N^2}\boldsymbol{l}\bfU_{\rm c}\,{\cal G}\right)+\int_{0}^{1}dt\,\frac{\wo'}{\lambda}\left(\dot\Psi,\pi_{01}\,\frac{\lambda^2} {N^2}\boldsymbol{m}\bfU_{\rm c}\,{\cal G}\right)\right),\\
\Delta_{\rm o}^{(1)}S_{\rm oc}=&N^2\left(\int_{0}^{1}dt\,\frac\wc{\lambda^2}\left(\dot\Phi,\pi_{10}\,\lambda\boldsymbol{l}\bfU^{'(1)}_{\rm o}\,{\cal G}\right)+\int_{0}^{1}dt\,\frac{\wo'}{\lambda}\left(\dot\Psi,\pi_{01}\,\lambda\boldsymbol{m}\bfU^{'(1)}_{\rm o}\,{\cal G}\right)\right),\\
\Delta_{\rm o}^{(2)}S_{\rm oc}=&N^2\left(\int_{0}^{1}dt\,\frac\wc{\lambda^2}\left(\dot\Phi,\pi_{10}\,\frac\lambda{N^2}\boldsymbol{l}\bfU^{'(2)}_{\rm o}\,{\cal G}\right)+\int_{0}^{1}dt\,\frac{\wo'}{\lambda}\left(\dot\Psi,\pi_{01}\,\frac\lambda{N^2}\boldsymbol{m}\bfU^{'(2)}_{\rm o}\,{\cal G}\right)\right).\\
\end{align}

The various $\bfU$ operators  are the coalgebra realizations of the open and closed Poisson bi-vectors \cite{quantum-OC}. In the closed sector we have
\begin{align}
U_{\rm c}=\frac{(-1)^{c^a}}{2}\,c_a\wedge c^a\quad\to\quad\bfU_{\rm c}=\frac{(-1)^{c^a}}{2}\,\boldsymbol{c}_a\boldsymbol{c}^a,
\end{align}
where $\boldsymbol{c}$ is the zero-product coderivation associated to the vector $c$. Structurally we see that the closed string laplacian $\Delta_{\rm c}$  corresponds to 
\begin{align}
\Delta_{\rm c}\to \frac{\lambda^2}{N^2}U_{\rm c}.
\end{align}
Similarly in the open string sector we have
\begin{align}
U_{\rm o}=\frac{(-1)^{o^a}}{2}\,o_a^{ij}\wedge o^a_{ij}\quad\to\quad\bfU_{\rm o}=\frac{(-1)^{o^a}}{2}\,\boldsymbol{o}_a^{ij}\boldsymbol{o}^a_{ij}.
\end{align}
While the action of $\bfU_{\rm c}$ on the open-closed tensor algebra is obvious, the action of $\bfU_{\rm o}$ is more involved. In particular if we act $\bfU_{\rm o}$ on the open-closed group-like element we get

\begin{equation}
\begin{split}
\label{ eqn: Uopen}
     \boldsymbol{U}_{\rm o}{\cal G} &=\dfrac{(-)^{o^{a}}}{2}\boldsymbol{o}^{a}_{ij}\boldsymbol{o}_{a}^{ij} \sum_{b,k,\{l_{1},...,l_{b}\}}^{\infty}\dfrac{1}{b!k!(l_{1})\cdot \cdot\cdot (l_{b})}\left(\Phi^{\wedge k} \otimes' \Psi^{\odot l_{1}}\wedge' \cdots\wedge'\Psi^{\odot l_{b}}\right)\\
     &=\dfrac{(-)^{o^{a}}}{2}\boldsymbol{o}^{a}_{ij} \sum_{b,k,\{l_{1},...,l_{b}\}}^{\infty}\dfrac{b(l_{1})}{b!k!(l_{1})\cdot \cdot\cdot (l_{b})}\left(\Phi^{\wedge k} \otimes' o_{a}^{ij}\odot\Psi^{\odot l_{1}}\wedge' \cdots\wedge'\Psi^{\odot l_{b}}\right)\\
     &=\dfrac{(-)^{o^{a}}}{2} \sum_{b,k,\{l_{1},...,l_{b}\}}^{\infty}\dfrac{b(l_{1})}{b!k!(l_{1})\cdot \cdot\cdot (l_{b})}\left(
        \Phi^{\wedge k}\otimes'\left[ \sum_{n=0}^{l_1} o^{a}_{ij}\odot\Psi^{\odot n} \odot o_{a}^{ij}\odot \Psi^{\odot l_{1}-n}   \right]\wedge' \cdots \wedge' \Psi^{\odot l_{b}}\right)\\
&+\dfrac{(-)^{o^{a}}}{2} \sum_{b,k,\{l_{1},...,l_{b}\}}^{\infty}\dfrac{b(b-1)(l_{1})(l_{2})}{b!k!(l_{1})\cdot \cdot\cdot (l_{b})}\left(\Phi^{\wedge k} \otimes' o_{a}^{ij}\odot\Psi^{\odot l_{1}}\wedge' o^{a}_{ij}\odot \Psi^{\odot l_{2}} \cdots\wedge'\Psi^{\odot l_{b}}\right)\\
&\coloneqq   \boldsymbol{U}_{\rm o}^{(1)}{\cal G} +\boldsymbol{U}_{\rm o}^{(2)}{\cal G}.
     \end{split}
\end{equation}
In other words
\begin{align}
 \boldsymbol{U}_{\rm o}^{(1)}=&\dfrac{(-)^{o^{a}}}{2}\boldsymbol{o}_{a}^{ij}\boldsymbol{o}^{a}_{ij}{\Bigg |}_{\textrm{same boundary}}\\
 \boldsymbol{U}_{\rm o}^{(2)}=&\dfrac{(-)^{o^{a}}}{2}\boldsymbol{o}_{a}^{ij}\boldsymbol{o}^{a}_{ij}{\Bigg |}_{\textrm{different boundaries}}.\\
 \end{align}
Again it is useful to define normalized versions (which have a finite large $N$ limit). Taking into account (\ref{ntrid1},\ref{ntrid2}) we see that they are given by
\begin{equation}
  \boldsymbol{U}_{o}'^{(1)}\coloneqq\dfrac{1}{N}\boldsymbol{U}_{o}^{(1)}  
\end{equation}
and 
\begin{equation}
\boldsymbol{U}_{o}'^{(2)}\coloneqq N\boldsymbol{U}_{o}^{(2)}.
\end{equation}
Finally we can write the relations with the open string laplacians as
\begin{align}
\Delta_{\rm o}^{(1)}\to \lambda \bfU_{\rm o}'^{(1)}
\end{align}
and
\begin{align}
\Delta_{\rm o}^{(2)}\to \frac\lambda{N^2}\bfU_{\rm o}'^{(2)}.
\end{align}
Now we can clearly see that in the limit $N\to\infty$ both $\Delta_{\rm c}$ and $\Delta_{\rm o}^{(2)}$ are suppressed as $N^{-2}$. This is in perfect agreement with the fact that the open-closed action \eqref{OC-WZWoc} is dominated by genus zero vertices in the large $N$ limit and therefore the BV structures that  increase the genus (which are precisely $\Delta_{\rm c}$ and $\Delta_{\rm o}^{(2)}$) should be suppressed as well.

\subsection{Planar homotopy relations and nilpotency}

If we now consider the planar action \eqref{OC-WZW-planar} we see that its master equation can be explicitly expressed as
\begin{align}
\frac12(S_{\rm pl},S_{\rm pl})_{\rm c}&+\frac12(S_{\rm pl},S_{\rm pl})_{\rm o}+\Delta_{\rm o}^{(1)}S_{\rm pl}\\
=&\int_{0}^{1}dt\,\frac\wc{\lambda^2}\left(\dot\Phi,\pi_{10}\,\left[\boldsymbol{l}^{(p)}\boldsymbol{l}^{(p)}+\boldsymbol{l}^{(p)}\boldsymbol{m}^{(p)}+\lambda\,\boldsymbol{l}^{(p)}\bfU_{\rm o}'^{(1)} \right]\,{\cal G}\right)\\
+&\int_{0}^{1}dt\,\frac{\wo'}{\lambda}\left(\dot\Psi,\pi_{01}\left[\boldsymbol{m}^{(p)}\boldsymbol{m}^{(p)}+\boldsymbol{m}^{(p)}\boldsymbol{l}^{(p)}+\lambda\,\boldsymbol{m}^{(p)}\bfU_{\rm o}'^{(1)}\right]\,{\cal G}\right).
\end{align}
Just as for the total open-closed action \cite{quantum-OC} we can repackage everything into a single structure (up to the constant $\Lambda_0$, \eqref{Lambda0})
\begin{align}
S_{\rm pl}(\Phi,\Psi)=\int_0^1\hw'\left(\dot\chi\,,\,\pi_1 \bfn^{(p)}\,{\cal G}\right),\label{Splanar-unit}
\end{align}
where
\begin{align}
\chi\coloneqq\Phi+\Psi&\,=(\pi_{10}+\pi_{01})\,{\cal G}=\pi_1\,{\cal G}\\
\hw'(\chi_1,\chi_2)\coloneqq&\,\frac{\wc}{\lambda^2}(\Phi_1,\Phi_2)+\frac{\wo'}{\lambda}(\Psi_1,\Psi_2)\\
 \bfn^{(p)}\coloneqq&\,\bfl^{(p)}+\bfm^{(p)}.
\end{align}
Then it is easy to realize that the planar BV master equation is solved provided
\begin{align}
(\bfn^{(p)}+\bfU^{(p)})^2=0,
\end{align}
where we have defined $\bfU^{(p)}\coloneqq \lambda\bfU_{\rm o}'^{(1)}$. Notice indeed that (upon use of the trivial vanishing of $\pi_1\bfU^{(p)}$ and $\left(\bfU^{(p)}\right)^2$) we have that
\begin{align}
\pi_{10}(\bfn^{(p)}+\bfU^{(p)})^2&=\pi_{10}\left[\boldsymbol{l}^{(p)}\boldsymbol{l}^{(p)}+\boldsymbol{l}^{(p)}\boldsymbol{m}^{(p)}+\lambda\,\boldsymbol{l}^{(p)}\bfU_{\rm o}'^{(1)} \right],\\
\pi_{01}(\bfn^{(p)}+\bfU^{(p)})^2&=\pi_{01}\left[\boldsymbol{m}^{(p)}\boldsymbol{m}^{(p)}+\boldsymbol{m}^{(p)}\boldsymbol{l}^{(p)}+\lambda\,\boldsymbol{m}^{(p)}\bfU_{\rm o}'^{(1)}\right].
\end{align}
This  homotopy structure (which was in fact already noticed  by Zwiebach years ago \cite{Zwiebach:1997fe},  without however an interpretation of its physical meaning) describes a (large $N$) theory where closed strings are purely classical  while open strings are fully quantum but restricted to the planar sector. It appears rather clear that this is a privileged starting point to understand microscopically how a large number of D-branes can back-react on the initial classical closed string geometry to a new classical closed string geometry, possibly without D-branes anymore. In the next section, purely driven by the consistency of the homotopy structure, we will attempt to make some  first step in this direction.
\section{Closed string backreaction}\label{sec:4.0}
The way to the closed string backreaction is composed of two steps, both of which require some assumption on the initial background. We first 
integrate out the open string degrees of freedom to end up with an intermediate closed string theory which (second step) has to be stabilized by canceling the tadpole induced by the world-sheet boundaries.

\subsection{Integrating out open strings}\label{sec:4}
After having constructed the planar open-closed action \eqref{Splanar-unit} we would like to integrate out the open string sector. Formally this can be done  
using the homotopy transfer \cite{Sen:2016qap, Jurco:2019yfd, Erbin:2020eyc, Koyama:2020qfb, Okawa:2022sjf } with a projector $P_{\rm c}=1_{{\cal H}_{\rm c}}\oplus 0_{{\cal H}_{\rm o}}$ 
\begin{align}
P_{\rm c}\,({\cal H}_{\rm c}\oplus{\cal H}_{\rm o})={\cal H}_{\rm c}.
\end{align}
Notice that obviously 
\begin{align}
[Q,P_{\rm c}]=0,
\end{align}
where $Q=Q_{\rm c }+Q_{\rm o}$.
To run the homotopy transfer we need a propagator $h_{\rm o}$ such that the Hodge-Kodaira decomposition is satisfied
\begin{align}
[Q,h_{\rm o}]=1-P_{\rm c}=P_{\rm o}=0_{{\cal H}_{\rm c}}\oplus 1_{{\cal H}_{\rm o}}.\label{H-K}
\end{align}
This equation may look surprising because $P_{\rm o}$ projects on the full open string Hilbert space where there is also the open string cohomology which evidently cannot be exact (in the form of $[Q,h_{\rm o}]$). Nevertheless this is the needed projection to integrate out the open string sector. 
Here we are  assuming to deal with a peculiar open-closed background such that the open string cohomology (if present) is prevented from propagating inside generic amplitudes with external closed strings. This is certainly a rather special situation that is not realized in generic D-brane systems. However there are cases in which the open string cohomology is really only present in the form of external states and the interactions of the theory are such that on-shell open strings are never produced inside a diagram. In the bosonic string this is explicitly realized for example in the minimal $(2,1)$ string with FZZT branes, as discussed extensively by Gaiotto and Rastelli \cite{Gaiotto:2003yb}, see also the discussion in the conclusions. A similar situation, although in the framework of the topological string, is expected to be realized in the Gopakumar-Vafa conifold transition \cite{Gopakumar:1998ki}.\footnote{In a slightly more generic situation we expect that the non-propagation of open string cohomology inside a diagram could be achieved by a suitable decoupling limit.}

 This is our working hypothesis without which it would not be possible to integrate out open strings completely, at least not in this simple way.\footnote{There could be instances in which the open string cohomology is present and propagating, but still it will have to be integrated out at the end, to end up with amplitudes with external closed strings only. This is realized for example in D-instanton contributions (see for example \cite{Sen:2019qqg}), in order to eventually obtain closed string momentum conservation.} Therefore we will just assume that our open-closed background allows for the possibility of integrating out open strings in the sense described above.  That said,
we can think of $h_{\rm o}$ as $\frac{b_0}{L_0}$ but we \ch{do not} necessarily want to commit to this choice.
Without loss of generality, we choose $h_{\rm o}$ so that $h_{\rm o}P_{\rm c}=P_{\rm c}h_{\rm o}=0$ as well as $h_{\rm o}^2=0$.
We also write
\begin{align}
\Pi_{\rm c}I_{\rm c}&=1_{{\cal H}_{\rm c}},\\
I_{\rm c}\Pi_{\rm c}&=P_{\rm c},
\end{align}
where $\Pi_{\rm c}$ and $I_{\rm c}$ are the canonical projection and inclusion between ${\cal H}_{\rm c}\oplus{\cal H}_{\rm o}$ and ${\cal H}_{\rm c}$. 
These objects define the strong deformation retract (see for example \cite{Erbin:2020eyc})
\begin{align}
\mathrel{\raisebox{+14.5pt}{\rotatebox{-110}{  \begin{tikzcd}[sep=12mm,
			arrow style=tikz,
			arrows=semithick,
			diagrams={>={Straight Barb}}
			]
			\ar[to path={ node[pos=.5,C] }]{}
			\end{tikzcd}
}}}\hspace{-2.7mm}
(-h_{\rm 0})\,(\mathcal{H}_{\rm c}\oplus\mathcal{H}_{\rm o},Q_{\rm c}+Q_{\rm o})
\hspace{-3.2mm}\raisebox{-0.2pt}{$\begin{array}{cc} {\text{\scriptsize $\Pi_{\rm c}$}}\\[-2.5mm] \mathrel{\begin{tikzpicture}[node distance=1cm]
		\node (A) at (0, 0) {};
		\node (B) at (1.5, 0) {};
		\draw[->, to path={-> (\tikztotarget)}]
		(A) edge (B);
		\end{tikzpicture}}\\[-3mm] {\begin{tikzpicture}[node distance=1cm]
		\node (A) at (1.5, 0) {};
		\node (B) at (0, 0) {};
		\draw[->, to path={-> (\tikztotarget)}]
		(A) edge (B);
		\end{tikzpicture}}\\[-3.5mm]
	\text{\scriptsize $I_{\rm c}$} \end{array}$} \hspace{-3.4mm}
(\mathcal{H}_{\rm c}, Q_{\rm c})\,.\label{eq:SDRH}
\end{align}
These operators can be upgraded to operators on the open-closed tensor algebra $e^{\wedge {\cal H}_{\rm c}}\otimes' e^{\wedge' {\cal CH}_{\rm o}}$  
\begin{align}
Q&\to\boldsymbol{Q}\\
P_{\rm c}&\to\boldsymbol{P}_{\rm c}\\
\Pi_{\rm c}&\to\boldsymbol{\Pi}_{\rm c}\\
I_{\rm c}&\to\mathbf{I}_{\rm c}\\
h_{\rm o}&\to\boldsymbol{h}_{\rm o},
\end{align} 
in such a way that 
\begin{align}
[\boldsymbol{Q},\boldsymbol{h}_{\rm o}]=&1-\boldsymbol{P}_{\rm c},\\
\boldsymbol{\Pi}_{\rm c}\mathbf{I}_{\rm c}=&1_{{\cal SH}_{\rm c}},\\
\mathbf{I}_{\rm c}\boldsymbol{\Pi}_{\rm c}=&\boldsymbol{P}_{\rm c}.
\end{align}
See appendix \ref{app:A} for the explicit construction. This allows to upgrade \eqref{eq:SDRH} to $e^{\wedge {\cal H}_{\rm c}}\otimes' e^{\wedge' {\cal CH}_{\rm o}}$  
\begin{align}\boldsymbol
\mathrel{\raisebox{+14.5pt}{\rotatebox{-110}{  \begin{tikzcd}[sep=12mm,
			arrow style=tikz,
			arrows=semithick,
			diagrams={>={Straight Barb}}
			]
			\ar[to path={ node[pos=.5,C] }]{}
			\end{tikzcd}
}}}\hspace{-2.7mm}
(-\boldsymbol{h}_{\rm 0})\,(e^{\wedge {\cal H}_{\rm c}}\otimes' e^{\wedge' {\cal CH}_{\rm o}},\boldsymbol{Q}_{\rm c}+\boldsymbol{Q}_{\rm o})
\hspace{-3.2mm}\raisebox{-0.2pt}{$\begin{array}{cc} {\text{\scriptsize $\boldsymbol{\Pi}_{\rm c}$}}\\[-2.5mm] \mathrel{\begin{tikzpicture}[node distance=1cm]
		\node (A) at (0, 0) {};
		\node (B) at (1.5, 0) {};
		\draw[->, to path={-> (\tikztotarget)}]
		(A) edge (B);
		\end{tikzpicture}}\\[-3mm] {\begin{tikzpicture}[node distance=1cm]
		\node (A) at (1.5, 0) {};
		\node (B) at (0, 0) {};
		\draw[->, to path={-> (\tikztotarget)}]
		(A) edge (B);
		\end{tikzpicture}}\\[-3.3mm]
	\text{\scriptsize $\mathbf{I}_{\rm c}$} \end{array}$} \hspace{-3.4mm}
(e^{\wedge {\cal H}_{\rm c}}, \boldsymbol{Q}_{\rm c})\,.\label{eq:SDRH2}
\end{align}
After this preparation, we can write the perturbed inclusions and projections as
\begin{align}
\mathbf{\tilde I}_{\rm c}\,&\coloneqq\,\frac{1}{1+\boldsymbol{h}_{\rm o}(\delta\bfn^{(p)}+\bfU^{(p)})}\,\mathbf{I}_{\rm c},\\
\boldsymbol{\tilde \Pi}_{\rm c}\,&\coloneqq\,\boldsymbol{ \Pi}_{\rm c}\,\frac{1}{1+(\delta\bfn^{(p)}+\bfU^{(p)})\boldsymbol{h}_{\rm o}},
\end{align}
where $\delta\bfn^{(p)}\coloneqq \bfn^{(p)}-\boldsymbol{Q}$ is the interacting part of the open-closed coderivation. Using well-known results, these objects obey
\begin{align}
\boldsymbol{\tilde \Pi}_{\rm c}\mathbf{\tilde I}_{\rm c}=\boldsymbol{ \Pi}_{\rm c}\mathbf{\tilde I}_{\rm c}=\boldsymbol{\tilde \Pi}_{\rm c}\mathbf{ I}_{\rm c}=\boldsymbol{\Pi}_{\rm c}\mathbf{I}_{\rm c}=1_{{\cal SH}_{\rm c}}.
\end{align}
We can use them to transfer the basic nilpotent structure $(\bfn^{(p)}+\bfU^{(p)})$ from ${\cal SH}_{\rm c}\otimes'{\cal SCH}_{\rm o}$ to ${\cal SH}_{\rm c}$
in the standard way
\begin{align}
\tilde\bfn^{(p)}=&\,\boldsymbol{ \Pi}_{\rm c}\,\bfn^{(p)}\mathbf{\tilde I}_{\rm c}=\boldsymbol{ \tilde\Pi}_{\rm c}\,\bfn^{(p)}\mathbf{ I}_{\rm c}=\boldsymbol{\tilde \Pi}_{\rm c}\,\bfn^{(p)}\mathbf{ \tilde I}_{\rm c}.\\
\tilde\bfU^{(p)}=&\,\boldsymbol{ \Pi}_{\rm c}\,\bfU^{(p)}\mathbf{\tilde I}_{\rm c}=\boldsymbol{ \tilde\Pi}_{\rm c}\,\bfU^{(p)}\mathbf{ I}_{\rm c}=\boldsymbol{\tilde \Pi}_{\rm c}\,\bfU^{(p)}\mathbf{ \tilde I}_{\rm c}.\\
\tilde{\boldsymbol{h}}_{\rm o}=&\,\mathbf{\tilde I}_{\rm c}\boldsymbol{h}_{\rm o}.
\end{align}
so that we can define the interacting strong deformation retract
\begin{align}\boldsymbol
\mathrel{\raisebox{+14.5pt}{\rotatebox{-110}{  \begin{tikzcd}[sep=12mm,
			arrow style=tikz,
			arrows=semithick,
			diagrams={>={Straight Barb}}
			]
			\ar[to path={ node[pos=.5,C] }]{}
			\end{tikzcd}
}}}\hspace{-2.7mm}
(-\tilde{\boldsymbol{h}}_{\rm 0})\,(e^{\wedge {\cal H}_{\rm c}}\otimes' e^{\wedge' {\cal CH}_{\rm o}},\bfn^{(p)}+\bfU^{(p)})
\hspace{-3.2mm}\raisebox{-0.2pt}{$\begin{array}{cc} {\text{\scriptsize $\tilde{\boldsymbol{\Pi}}_{\rm c}$}}\\[-2.5mm] \mathrel{\begin{tikzpicture}[node distance=1cm]
		\node (A) at (0, 0) {};
		\node (B) at (1.5, 0) {};
		\draw[->, to path={-> (\tikztotarget)}]
		(A) edge (B);
		\end{tikzpicture}}\\[-3mm] {\begin{tikzpicture}[node distance=1cm]
		\node (A) at (1.5, 0) {};
		\node (B) at (0, 0) {};
		\draw[->, to path={-> (\tikztotarget)}]
		(A) edge (B);
		\end{tikzpicture}}\\[-2.5mm]
	\text{\scriptsize $\tilde{\mathbf{I}}_{\rm c}$} \end{array}$} \hspace{-3.4mm}
(e^{\wedge {\cal H}_{\rm c}}, \tilde\bfn^{(p)}+\tilde\bfU^{(p)})\,.\label{eq:SDRH3}
\end{align}
Both transferred objects $\tilde\bfn^{(p)}$ and $\tilde\bfU^{(p)}$ are however peculiar because of the nature of the projection which kills all open string degrees of freedom.
In particular we immediately notice that
\begin{align}
\tilde\bfU^{(p)}\equiv 0,
\end{align}
because $\bfU^{(p)}$ creates open string states which are killed by the projection or the inclusion. Moreover we also realize that the transfer of $\bfn^{(p)}$ is actually a closed string coderivation and it is in fact the transfer of the closed string coderivation $\bfl^{(p)}$
\begin{align}
\tilde\bfn^{(p)}=\tilde\bfl=\boldsymbol{ \Pi}_{\rm c}\,\bfl^{(p)}\mathbf{\tilde I}_{\rm c}.
\end{align}
All of these results are summarized by the chain homotopy relations
\begin{align}
\mathbf{\tilde I}_{\rm c}\tilde\bfl=&\bfn^{(p)}\mathbf{\tilde I}_{\rm c}\\
\tilde\bfl\,\boldsymbol{\tilde \Pi}_{\rm c}=&\boldsymbol{\tilde \Pi}_{\rm c}\bfn^{(p)},
\end{align}
as well as
\begin{align}
0=&\,\bfU^{(p)}\mathbf{\tilde I}_{\rm c}\\
0=&\,\boldsymbol{\tilde \Pi}_{\rm c}\bfU^{(p)},
\end{align}
from which we can easily obtain that $\tilde \bfl^{(p)}$ is actually a nilpotent coderivation
\begin{align}
\tilde \bfl\,\tilde \bfl=\tilde \bfl\, (\boldsymbol{\tilde \Pi}_{\rm c}\mathbf{\tilde I}_{\rm c})\,\tilde\bfl=\boldsymbol{\tilde \Pi}_{\rm c}\,\bfn^{(p)}\bfn^{(p)}\,\mathbf{\tilde I}_{\rm c}=-\boldsymbol{\tilde \Pi}_{\rm c}\,(\bfn^{(p)}\bfU^{(p)}+\bfU^{(p)}\bfn^{(p)})\,\mathbf{\tilde I}_{\rm c}=0,
\end{align}
where we used the original planar open-closed homotopy algebra
\begin{align}
(\bfn^{(p)}+\bfU^{(p)})^2=\bfn^{(p)}\bfn^{(p)}+\bfn^{(p)}\bfU^{(p)}+\bfU^{(p)}\bfn^{(p)}=0.
\end{align}
This means that $\tilde\bfl$ describes an $L_\infty$ algebra.  Since the homotopy transfer performs the path integral on $\Psi$  with the gauge fixing $h_{\rm o}\Psi=0$, the  effective action constructed from $\tilde \bfl$  is  given by
\begin{align}
\int {\cal D}\Psi &\,e^{-S_{\rm pl}(\Phi,\Psi)}{\Big |}_{h_{\rm o}\Psi=0}=e^{-S_{\rm eff}(\Phi)}\\
S_{\rm eff}(\Phi)=&\int_0^1dt\,\frac{\wc}{\lambda^2}\left(\dot\Phi\,,\,\pi_1\tilde\bfl\,e^{\wedge\Phi}\right)+\Lambda_{\rm open} +\Lambda_0,\label{eff-action}
\end{align}
where the constant part of the action $\Lambda_{\rm open}$ is generated by the path integral from the kinetic term and the zero products. The effective action \eqref{eff-action} is structurally the same as \eqref{S-planar}, with all the open strings set to zero (but not the boundaries)
\begin{align}
S_{\rm eff}(\Phi)=&\,\sum_{b=0}^{\infty} \lambda^{b-2}\sum_{k=0}^{\infty}\frac{1}{b!k!}\mathcal{A''}^{g=0,b}_{k}\left(\Phi^{\wedge k}\right),\\
 \mathcal{A''}^{g=0,b}_{k+1}(\Phi_1\wedge\cdots\wedge\Phi_{k+1})=&\,\wc\left(\Phi_1,\tilde l_{k}^{\ch{(b)}}\left(\Phi_2\wedge\cdots\wedge \Phi_{k+1}\right)\right),\\
 \tilde l_{k}=&\,\pi_1\tilde\bfl\pi_k \ch{=\sum_{b=0}^\infty \lambda^b\,\tilde l_k^{(b)}}\,.
\end{align}
The difference is that now the off-shell amplitudes $\mathcal{A''}^{g=0,b}_{k}$ are integrated in moduli space also in the regions of open string degeneration, thanks to the integration-out of the open strings that we have just performed. The remaining part of integration towards closed string degeneration will be performed by the Feynman diagrams of the obtained effective action, through the closed string propagators.

\subsection{Eliminating the closed string tadpole}\label{sec:5}
Notice that the  effective closed string theory we have just obtained is based on a {\it weak} $L_\infty$-algebra. Indeed there is clearly a tadpole represented by a genus zero surface with a closed string puncture and an arbitrary number of boundaries weighted by the 't Hooft coupling. In particular, in a specific choice of SFT data we can write
\begin{align}
\tilde l_0=\lambda e^{-sL_0^+}\,\ket {B_0'}+O(\lambda^2),
\end{align}
where $\ket  {B_0'}$  is the initial (normalized) boundary state. The full tadpole can be written as a power series expansion in the 't Hooft coupling as
\begin{align}
\tilde l_0=\sum_{b=1}^\infty \lambda^b\,\tilde l_0^{(b)}.
\end{align}
Notice that there is no $b=0$ contribution (the original bulk CFT \ch{is assumed to have} no \ch{sphere} tadpole).
In fact the same expansion in boundaries can be done for all other closed strings products
\begin{align}
\tilde l_k=\sum_{b=0}^\infty \lambda^b\,\tilde l_k^{(b)},
\end{align}
where this time also the sphere $b=0$ contributions are contained.
Because of the tadpole we cannot say that we have obtained a classical closed string background. To achieve this we have to solve the tadpole-sourced closed string field theory equation of motion\footnote{A very similar equation has been considered in \cite{Khoury:1999szo} as a way to regulate the collisions of the deformed world-sheet  sigma model with boundaries.}
\begin{align}
\sum_{k=1}^\infty\,\frac1{k!}\tilde l_{k}(\Phi^{\wedge k})&=-\tilde l_0,\label{vacuum-eom}
\end{align}
which, if expanded in terms of the 't Hooft coupling,  gives rise to the recursive equations
\begin{align}
\Phi=&\sum_{n=1}^\infty\,\lambda^n\, \Phi_n,\\
O(\lambda):\quad& Q_{\rm c}\Phi_1=-\tilde l_0^{(1)}\label{QCB}\\
O(\lambda^2):\quad&  Q_{\rm c}\Phi_2=-\frac12 \tilde l_2^{(0)}(\Phi_1^{\wedge2})-\tilde l_1^{(1)}(\Phi_1)-\tilde l_0^{(2)}\\
O(\lambda^3):\quad&  Q_{\rm c}\Phi_3=-\frac16 \tilde l_3^{(0)}(\Phi_1^{\wedge3})- \tilde l_2^{(0)}(\Phi_1\wedge\Phi_2)-\frac12\tilde l_2^{(1)}(\Phi_1^{\wedge2})-\tilde l_1^{(1)}(\Phi_2)-\tilde l_1^{(2)}(\Phi_1)-\tilde l_0^{(3)}\\
\vdots&\0\quad\quad\quad.
\end{align}
The $O(\lambda)$ equation that we found has been already discussed in the past. For example (implicitly) in \cite{DiVecchia:1997vef}, to extract the large distance behaviour of the supergravity $p$-brane metric from the boundary state, or  in \cite{Sen:2004zm} to read-off the emitted closed string radiation from the boundary state of a decaying D-brane. 

This  equation is rather subtle: indeed already at $O(\lambda)$, \eqref{QCB} is apparently obstructed since $\tilde l_0^{(1)}\sim\ket B$ is clearly in the closed string cohomology and therefore it cannot be equal to $Q\Psi_1$ if $\Psi_1$ has to be a regular string field. However the obstruction is in the zero momentum sector and if the boundary state has enough Dirichlet directions, the obstruction will be contained in the continuous momentum integral of the boundary state and therefore can be avoided by continuity.  This is in fact fully analogous to how the Coulomb potential is formed as a vacuum shift solution of electromagnetism in presence of a delta-function source and the meaning of the obstruction is essentially that there is no Coulomb potential if the dimensions transverse to the charge are compact, so that the zero momentum part of the source would give an isolated equation without solution. On the other hand, the higher order $\lambda$-corrections seem  to mark an important difference from the simple backreaction of a source, which is a solution with a  singularity at the position of the source itself. Indeed we expect that if the full vacuum shift solution exists it \ch{should not} be singular at the locus of the initial D-branes. In this sense perhaps an expansion in $\lambda$ is not the best way to proceed, since every power of $\lambda$ may be singular but the whole resummation may be regular\footnote{Think for example about the regular ``instanton'' $\frac{\lambda}{x^2+\lambda^2}$, when it is expanded in $\lambda=0$.}. Understanding the structure of this equation and its possible solution seems a rather fundamental problem that is universal in every situation in which a large number of D-branes backreacts on the initial closed-string background. 

Suppose however that we have found a solution $\Phi_*(\lambda)$ to \eqref{vacuum-eom}. Then this will give rise to a new closed string background without a tadpole anymore
\begin{align}
S(\varphi)\coloneqq S_{\rm eff}(\Phi_*(\lambda)+\varphi)=\int_0^1dt\,\frac{\wc}{\lambda^2}\left(\dot\varphi\,,\,\pi_1\tilde\bfl_*\,e^{\wedge\varphi}\right)+\Lambda_{\rm open} +\Lambda_0+ \Lambda_{\rm closed},
\end{align}
where the shifted coderivation $\tilde\bfl_*$ now describes a strong $L_\infty$ structure 
\begin{align}
[\tilde\bfl_*,\tilde\bfl_*]&=0\\
\tilde\bfl_*\pi_0&=0.
\end{align}
Notice that the closed string shift generates a new piece of vacuum energy so that in total we now have
\begin{align}
\Lambda_{\rm tot}=\Lambda_{\rm open} +\Lambda_0+ \Lambda_{\rm closed}=\sum_{b=1}^\infty \,\lambda^{b-2} \Lambda_b.
\end{align}
The meaning of this full cosmological constant should be the genus zero partition function summed over all possible boundaries, with a full moduli space integration. In particular $\Lambda_0$ integrates in the middle of moduli space, $\Lambda_{\rm open}$ (generated by the open string integration-out) takes care of the open-string degeneration regions and finally $\Lambda_{\rm closed}$ (generated by the closed string vacuum shift) integrates towards closed string degenerations. It would be interesting to study this vacuum energy more concretely.

The next term in the obtained stable closed string action is the kinetic term
\begin{align}
S_2(\varphi)=\frac1{2\lambda^2}\wc\left(\varphi,\tilde l^*_1\varphi\right),
\end{align}
where 
\begin{align}
\tilde l^*_1=\pi_1\tilde\bfl_*\pi_1=Q_{\rm c}+\sum_{b=1}^\infty\,\lambda^b\, \tilde l^{*(b)}_1.
\end{align}
The cohomology of this operator can be studied in an expansion in $\lambda$ for example following  section 7 of \cite{Sen:2019jpm} (see also appendix B of \cite{cosmo}) and this is expected to give the physical closed string spectrum on the `dual' closed string geometry. Continuing in this way we can describe the new closed string background in powers of $\lambda$. It would be clearly very instructive to test this in some explicit and tractable example.
\section{Discussion}\label{sec:6}
Although the construction of this paper is based on the bosonic string, the homotopy structure we have discussed is expected to be realized in the superstring as well (although it has not yet been explicitly constructed, see \cite{Kunitomo:2022qqp, super-SDHA} for recent progress). Therefore our picture is expected to be universal for any string theory containing D-branes. This includes theories that are  consistent at the quantum level like the $c=1$ and the \ch{minimal} string theories (which are just specific examples of bosonic strings), the topological string and of course Type II A/B superstring. By generalizing our construction to non-orientable world-sheets \cite{Moosavian:2019ydz}, we could also treat along the same lines the $SO(32)$ Type I superstring. 

A main outcome of this paper is the characterization of the geometric transition as the integration-out of open strings, followed by a vacuum-shift for the closed strings. We have seen that integrating out the open strings results in a closed string field theory that is still defined on the initial un-backreacted background where however there is a tadpole.  This classical closed string theory is just  the initial pure closed string field theory without D-branes with the addition of a deformation consisting of closed strings off-shell amplitudes on genus zero surfaces with boundaries. This is a huge deformation of the initial closed string background which is controlled by the 't Hooft coupling $\lambda=\kappa N$, a continuous tunable parameter. Although it seems that for $\lambda=0$ we have just  the initial purely closed string field theory without D-branes, the limit $\lambda\to0$ may not in general be continuous because D-branes cannot  be  continuously ``switched-off''.  This seems to mark a  difference with the situation analyzed in \cite{cosmo}, where instead we were interested in how the open string dynamics is affected by closed strings deformations, which can be instead switched-off continuously.  
It seems reasonable that the possibility of integrating out completely the open string sector and the existence of the vacuum shift solution are related, because singularities at the locus of the source are expected to have their origin in divergences from open string degenerations \cite{Douglas:1996yp}, which in turn are associated with open string obstructions. More explorations will be needed on this front.
Still driven by our algebraic considerations, instead of integrating out the open strings, we can also study the problem of integrating out the closed strings, to end up, if the given background allows for this, with a pure quantum open string field theory. It would be interesting to explore this other direction of the integration-out process, together with its expected subtleties. All in all, the open-closed SFT approach we are proposing allows to see the two sides of the open-closed duality simply as two aspects of the same quantum field theory. Our considerations in this paper, however,  are only based on the algebraic BV-structure of open-closed SFT in the large $N$ limit and  the need of some explicit example to test our construction is compelling.

A rather clear example where the backreaction we describe in this paper is smoothly realized starts by considering  $N$ FZZT branes \cite{Fateev:2000ik} with open string moduli  given by the boundary cosmological constant(s) in the the initial background of the $(2,1)$ minimal string theory \cite{DiFrancesco:1993cyw}. In this case, following the work of \cite{Gaiotto:2003yb},  the backreaction of the FZZT branes can indeed be completely absorbed in a change in the closed string background which, depending on the initial values of the boundary cosmological constants, will move in the continuous manifold containing the $(2,2k+1)$ minimal strings. In this case the absence of propagating open string cohomology is built in the theory and our assumptions on open string integration-out are realized explicitly.  We are working at the moment  on the details of this duality in the context of our open-closed SFT setting \cite{minimal OC}.  An analogous realization should be given by the original conifold transition \cite{Gopakumar:1998ki, Ooguri:2002gx}, although in this case we should appropriately depart from the simple bosonic string setting we have been assuming in this paper.  
These are  very peculiar examples in which the D-branes can be completely absorbed in a new closed string background. However these paradigms of open-closed duality are not expected to be representatives of the generic situation in which a large number of D-branes backreacts on a given closed string background. In general we expect that the source may remain after the backreaction and a decoupling limit of some kind (depending on the specific example under consideration) will be needed to end up with a pure closed string background. This is for example the standard understanding of how closed strings on $AdS_5\times S^5$ emerge as the backreaction of a large number of D3-branes in flat space, only after taking the low-energy limit $\alpha'\to0$, which decouples the initial open and closed strings \cite{Maldacena:1997re}. The need of some kind of decoupling limit seems to be a rather generic situation which we expect to manifest itself in obstructions in integrating out completely the open string sector in our open-closed SFT approach. Possible obstructions in solving the closed string vacuum shift equations should also be expected and carefully studied.  We hope to be able to understand these important issues in the near future.

Let's end with some further speculation. The typical setting of open/closed duality is based on the two `external' parameters  $N$ and $\lambda$. However in open-closed string field theory these may not be free parameters after all.

As far as $N$ is concerned we know that in Witten OSFT it is possible to build universal solutions for generic $N$ identical D-branes, starting from the OSFT defined on a single D-brane \cite{Erler:2019fye, Erler:2014eqa} and we expect that these saddle points should remain in the full open-closed SFT, at least in the bosonic string.  Therefore the 't Hooft limit may be thought of as concentrating on a region of the full open-closed field space near the saddle corresponding to a large number of D-branes. In this regard we expect that a somewhat special role should be played by $N=0$ which, at least in the present bosonic setting, should be understood as having the open string sector at the tachyon vacuum  \cite{Sen:1999xm, Schnabl:2005gv}. In this case we would expect that integrating out the (perturbatively) trivial open string sector should not change the starting closed string background. It would be interesting to be able to do this operation exactly and fully characterize the final closed string background to verify that indeed there is no backreaction.

The string coupling constant $\kappa$ and consequently the  't Hooft parameter $\lambda$  are also not expected to be free parameters when closed strings are dynamical. In pure (critical) closed string field theory we know that $\kappa$ can be changed by giving a vacuum expectation value to the ghost dilaton \cite{Bergman:1994qq, Belopolsky:1995vi, Erler:2022agw} and we expect that this persists in the full open-closed string field theory \cite{progress1}. We could then explore different regimes of $\lambda$ by changing $\kappa$ through ghost-dilaton deformations.

Working in the complete framework of open-closed SFT also allows to connect with Okawa's idea \cite{Okawa:2020llq} of using Witten-type open (super)string field theory with Ellwood invariants \cite{Hashimoto:2001sm, Gaiotto:2001ji, Ellwood:2008jh}, to characterize the dual or backreacted closed string theory. Starting from the full open-closed SFT, before integrating out the open strings, we can integrate out all unphysical closed strings, projecting on the closed string cohomology. This gives an open string field theory deformed by couplings with on-shell closed strings. In this implementation, one can consider the singular limit of interpolating string field theories \cite{Zwiebach:1992bw} (see also \cite{Cho:2019anu}) to end up precisely with Witten theory deformed by Ellwood invariants.\footnote{We expect this to be plainly doable for generic closed strings which are $(0,0)$ primaries. The ghost dilaton is a distinguished exception which may play a non-trivial role. To our knowledge and attempts from \cite{Maccaferri:2021ksp, Maccaferri:2021lau, Maccaferri:2021ulf}, we don't know how to generalize the Ellwood invariant to couple the ghost dilaton to Witten-type OSFT's.}  The correlation functions of these Ellwood invariants will be the same as the closed string on-shell amplitudes computed in the weak $L_\infty$ theory, with the difference that all the moduli space will now be covered by open string propagators, also in the regions of closed string degeneration. Therefore integrating out the open strings in Witten theory also performs, in a sense, the closed string vacuum shift, but then closed strings are constrained to be on-shell (with respect to the initial background). It would be interesting to see how, in this singular limit,  the data of the emerging closed string background could be somehow encapsulated in the open string dynamics.

We hope we will be able to understand these and related challenging points in the future.

\section*{Acknowledgments}
 We thank Ted Erler, Raghu Mahajan and Yuji Okawa for discussions. We thank the organizers and the participants of the workshop ``Matrix Models and String Field Theory", Benasque, May 2023 for the very stimulating atmosphere where this work has been presented. JV thanks INFN Turin for their hospitality during the initial stages of this work. The work of CM  and AR  is partially supported by the MUR PRIN contract 2020KR4KN2  String Theory as a bridge between Gauge Theories and Quantum Gravity and by the INFN project ST$\&$FI String Theory and Fundamental Interactions.    The work of JV was supported by the NCCR SwissMAP that
    is funded by the Swiss National Science Foundation.

\appendix

\section{Action of the propagator on \texorpdfstring{$\mathcal{SH}_{\rm c}\otimes'\mathcal{SCH}_{\rm o}$}{TEXT} }\label{app:A}

In this appendix we want to promote a generic propagator $h$ defined on ${\cal H}_{\rm c}\oplus {\cal H}_{\rm o}$ to $\mathcal{SH}_{c}\otimes'\mathcal{SCH}_{o}$ and prove the validity of the Hodge-Kodaira decomposition in that case. To start with, we \ch{recall} the well-known definitions of the propagator $\boldsymbol{h}_{TH}:\mathcal{TH}\longrightarrow \mathcal{TH}$ and $\boldsymbol{h}_{SH}:\mathcal{SH}\longrightarrow \mathcal{SH}$, see for example \cite{Erbin:2020eyc}
\begin{equation}
\boldsymbol{h}_{TH}\pi_{n}=\sum_{i=0}^{n-1}\left(1^{\otimes i}\otimes h \otimes P^{\otimes n-i-1}\right)\pi_{n},
\end{equation}
\begin{equation}
\boldsymbol{h}_{SH}\pi_{n}=\dfrac{1}{n!}\sum_{i=0}^{n-1}\left(1^{\wedge i}\wedge h \wedge P^{\wedge n-i-1}\right)\pi_{n},
\end{equation}
which satisfy the corresponding Hodge-Kodaira decomposition
\begin{equation}
\left[\boldsymbol{h}_{TH},\boldsymbol{Q}_{TH}\right]=1_{TH}-\boldsymbol{P}_{TH},
\end{equation}
\begin{equation}
\left[\boldsymbol{h}_{SH},\boldsymbol{Q}_{SH}\right]=1_{SH}-\boldsymbol{P}_{SH},
\end{equation}
where $\boldsymbol{Q}$ and $\boldsymbol{P}$ are defined as usual i.e., as coderivation and cohomomorphism respectively on the reference space
\begin{align}
&\boldsymbol{P}_{TH}\pi_{n}=P^{\otimes n}\pi_{n},\\
&\boldsymbol{P}_{SH}\pi_{n}=\dfrac{1}{n!}P^{\wedge n}\pi_{n},\\
&\boldsymbol{Q}_{TH}\pi_{n}=\sum_{i=0}^{n-1} \left(1^{\otimes i}\otimes Q \otimes 1^{\otimes n-i-1}\right)\pi_{n},\\
&\boldsymbol{Q}_{SH}\pi_{n}=\dfrac{1}{n!}\sum_{i=0}^{n-1} \left(1^{\wedge i}\wedge Q \wedge 1^{\wedge n-i-1}\right)\pi_{n}.
\end{align}
Quite naturally we extend these definitions to the cyclic tensor algebra  $\mathcal{CH}$
\begin{align}
 &\boldsymbol{h}_{CH}\pi_{n}=\dfrac{1}{n}\sum_{i=0}^{n-1}\left(1^{\odot i}\odot h \odot P^{\odot n-i-1}\right)\pi_{n},\\
 &\boldsymbol{P}_{CH}\pi_{n}=\dfrac{1}{n}P^{\odot n}\pi_{n},\\
 &\boldsymbol{Q}_{CH}\pi_{n}=\dfrac{1}{n}\sum_{i=0}^{n-1} \left(1^{\odot i}\odot Q \odot 1^{\otimes n-i-1}\right)\pi_{n}.
 \end{align}
We then proceed by demonstrating the validity of the Hodge-Kodaria decomposition 
\begin{equation}
\begin{split}
\boldsymbol{Q}_{CH}\boldsymbol{h}_{CH}\pi_{n}&=\dfrac{1}{n}\sum_{i=0}^{n-1}\boldsymbol{Q}_{CH}\left(1^{\odot i}\odot h \odot P^{\odot n-i-1}\right)\pi_{n}\\
&=\dfrac{1}{n}\sum_{i=0}^{n-1}\biggl\{\sum_{j=0}^{i-1}\left(1^{\odot j}\odot Q \odot 1^{\odot i-j-1}\odot h \odot P^{\odot n-i-1}\right)+\\
&\qquad\qquad+ 1^{\odot i}\odot Qh \odot P^{\odot n-i-1}+\\
&\qquad\qquad-\sum_{j=0}^{n-i-2}1^{\odot i}\odot h\odot P^{\odot j} \odot PQ \odot P^{\odot n-i-j-2}\biggr\}\pi_{n},
\end{split}
\end{equation}
\begin{equation}
\begin{split}
\boldsymbol{h}_{CH}\boldsymbol{Q}_{CH}\pi_{n}&=\dfrac{1}{n}\sum_{i=0}^{n-1}\left(1^{\odot i}\odot h \odot P^{\odot n-i-1}\right)\boldsymbol{Q}_{CH}\pi_{n}\\
&=\dfrac{1}{n}\sum_{i=0}^{n-1}\biggl\{-\sum_{j=0}^{i-1}\left(1^{\odot j}\odot Q \odot 1^{\odot i-j-1}\odot h \odot P^{\odot n-i-1}\right)+\\
&\qquad\qquad+ 1^{\odot i}\odot Qh \odot P^{\odot n-i-1}+\\
&\qquad\qquad+\sum_{j=0}^{n-i-2}1^{\odot i}\odot h\odot P^{\odot j} \odot PQ \odot P^{\odot n-i-j-2}\biggr\}\pi_{n},
\end{split}
\end{equation}
so that we get
\begin{equation}
\begin{split}
\left[\boldsymbol{Q}_{CH},\boldsymbol{h}_{CH}\right]\pi_{n}&=\dfrac{1}{n}\sum_{i=0}^{n-1}\left(1^{\odot i}\odot \left[Q,h\right]\odot P^{\odot n-i-1}\right)\pi_{n}\\
&=\dfrac{1}{n}\sum_{i=0}^{n-1}\left(1^{\odot i+1}\odot P^{\odot n-i-1}-1^{\odot i}\odot P^{\odot n-i}\right)\pi_{n}\\
&=\dfrac{1}{n}\left(1^{\odot n}-P^{\odot n}\right)\pi_{n}=\left(1_{CH}-\boldsymbol{P}_{CH}\right)\pi_{n}.
\end{split}
\end{equation}
Now having defined such operators we can easily extend them when applied to spaces obtained by tensoring $\mathcal{CH}$,$\mathcal{SH}$ and/or $\mathcal{TH}$.  For example let us consider $\mathcal{SCH}$
\begin{align}
    &\boldsymbol{P}_{SCH}\hat{\pi}_{b}=\dfrac{1}{b!}\boldsymbol{P}_{CH}^{\wedge' b}\hat{\pi}_{b},\\
&\boldsymbol{Q}_{SCH}\hat{\pi}_{b}=\dfrac{1}{b!}\sum_{i=0}^{b-1} \left(1_{CH}^{\wedge' i} \wedge'\boldsymbol{Q}_{CH} \wedge' 1_{CH}^{\wedge' b-i-1}\right)\hat{\pi}_{b},\\
&\boldsymbol{h}_{SCH}\hat{\pi}_{b}=\dfrac{1}{b!}\sum_{i=0}^{b-1}\left(1_{CH}^{\wedge' i}\wedge' \boldsymbol{h}_{CH} \wedge' \boldsymbol{P}_{CH}^{\wedge' b-i-1}\right)\hat{\pi}_{b},
\end{align}
where $\hat{\pi}_{b}$ is the projector that selects the terms in which there are  $b$ $\wedge'$-products of $\mathcal{CH}$ (or in other words all possible states to be inserted in $b$ boundaries). Following the same steps illustrated above, it can be shown that the Hodge-Kodaira decomposition is also satisfied here
\begin{equation}
\left[\boldsymbol{h}_{SCH},\boldsymbol{Q}_{SCH}\right]=1_{SCH}-\boldsymbol{P}_{SCH}.
\end{equation}
Finally, it is easy to treat the case of our interest i.e. the defintion of the Hodge-Kodaira operators on $\mathcal{SH}_{\rm c}\otimes' \mathcal{SCH}_{\rm o}$
\begin{align}
&\boldsymbol{P}=\boldsymbol{P}_{SH}\otimes' \boldsymbol{P}_{SCH},\\
&\boldsymbol{Q}=\boldsymbol{Q}_{SH}\otimes'1_{SCH}+1_{SH}\otimes' \boldsymbol{Q}_{SCH},\\
&\boldsymbol{h}=\boldsymbol{h}_{SH}\otimes' \boldsymbol{P}_{SCH}+1_{SH}\otimes'\boldsymbol{h}_{SCH},
\end{align}
and obviously it is true that 
\begin{equation}
\left[\boldsymbol{Q},\boldsymbol{h}\right]=1_{SH\otimes'SCH}-\boldsymbol{P}.
\end{equation}
Notice that these operators simplify when we choose $P=P_{\rm c}$ as in the main text, to integrate out completely the open string sector, because in this case every entry  of the open string tensor algebra (except the \ch{zeroth} tensor power) is killed by $P$.

\end{document}